# Using Early Quality Assurance Metrics to Focus Testing Activities

*Frank Elberzhager[1], Jürgen Münch[2]*

[1]Fraunhofer IESE, [2]University of Helsinki

[1]frank.elberzhager@iese.fraunhofer.de, [2]juergen.muench@cs.helsinki.fi

*Abstract:*

*Testing of software or software-based systems and services is considered as one of the most effort-consuming activities in the lifecycle. This applies especially to those domains where highly iterative development and continuous integration cannot be applied. Several approaches have been proposed to use measurement as a means to improve test effectiveness and efficiency. Most of them rely on using product data, historical data, or in-process data that is not related to quality assurance activities. Very few approaches use data from early quality assurance activities such as inspection data in order to focus testing activities and thereby reduce test effort. This article gives an overview of potential benefits of using data from early defect detection activities, potentially in addition to other data, in order to focus testing activities. In addition, the article sketches an integrated inspection and testing process and its evaluation in the context of two case studies. Taking the study limitations into account, the results show an overall reduction of testing effort by up to 34%, which mirrors an efficiency improvement of up to about 50% for testing.*

## 1    Introduction

Software and software-intensive systems are part of everyone's life and can be found all around us. Moreover, the size and complexity of such systems are continuously growing. Charette [1], for instance, states that in 2005, a typical cellphone contained about two million lines of code; nowadays, such phones may contain ten times as many. Another example are modern cars, with an estimate of 100 million lines of code. Consequently, developing high-quality software is becoming more challenging and expensive.

Jackson et al. [2] state that due to "the growth in complexity and invasiveness of software systems, the risk of a major catastrophe in which software failure plays a part is increasing." Boehm and Basili [3] mention that between 40 and 50 percent of all delivered software contains non-trivial defects. Humphrey [4] confirms that "today's large-scale systems typically have many defects".

Hence, in order to ensure software products of high quality, quality assurance (QA) activities play a crucial role in software development today. As far as analytical QA is concerned, a lot of well-established static and dynamic QA activities and techniques exist, such as inspections and testing [5-7]. However, while costs can increase dramatically when certain defects (especially critical ones) are not found, costs for conducting QA activities can also be a major cost driver during software development. This holds especially for testing activities. Myers [8] already stated that testing can consume approximately 50% of the development time and more than 50% of the overall development costs, which has been confirmed by recent studies [9-11].

Therefore, improving defect detection rates and reducing the costs for QA in general and for testing in particular are nowadays two of the major challenges (and goals) when conducting QA. Goals such as reduced effort or improved effectiveness are, for instance, addressed by automation and tool usage (e.g., [12-13]), defect prediction approaches (e.g., reliability growth models [14]), or by approaches predicting defect-prone parts in order to focus QA activities (e.g., [15], [17]). Goals that should be achieved in a concrete context may result in a concrete QA strategy.

With respect to approaches that allow focusing testing activities, specific product metrics are usually applied, such as size or complexity (e.g., [15-17]), or historical defect data are considered [18]. However, defect data from early QA activities, such as inspections, are often not used for focusing testing activities, and synergy effects between inspections and testing are often not exploited. Hence, an integrated inspection and testing approach called In$^2$Test was developed for predicting defect-prone parts and defect types for testing based on different inspection metrics. In order to be able to perform such predictions, knowledge about the relationships between inspections and testing is necessary. If such knowledge is not available, assumptions have to be made, which will again require validation in subsequent QA runs.

An evaluation of the In$^2$Test approach in two case studies showed that effort for testing can be reduced by up to 34% while maintaining a comparable quality level in the given environments. The actual effort reduction depends on the assumptions made. Moreover, improved knowledge about relationships between inspections and testing in the given contexts could lead to improved QA, and forms the basis for future research on the one hand, while offering explicit support for practitioners to improve their QA on the other hand.

This article presents the integrated inspection and testing approach In$^2$Test, emphasizes different QA improvement goals that can be achieved with the integrated approach, summarizes the main evaluation results of two case studies and discusses consequences, and indicates future research directions.

The remainder of this article is structured as follows: Section 2 presents the integrated inspection and testing approach, different benefits, and how they can be achieved by the integrated approach. The basic results from two case studies are summarized in Section 3, together with implications for practitioners. Finally, Section 4 concludes the article and gives an outlook on future work.

## 2 Approach

The main idea of the <u>integrated inspection</u> and <u>testing</u> approach In²Test is to use inspection metrics to focus testing activities. In order to ensure that the inspection data is valid, quality monitoring is performed before testing is focused based on the inspection data. If the quality of the inspection defect data is sufficient, different metrics can be applied, for instance, defect content (i.e., absolute number of inspection defects) or defect density (i.e., number of inspection defects per unit (e.g., lines of code)).

Furthermore, in order to be able to prioritize, you must know about the relationships between inspections and testing. Otherwise, assumptions must be made, such as that more defects are expected to be found during testing in those parts of the code where a significant number of defects are found during inspection. Such an assumption can be used to focus testing activities on code classes with the highest defect content or the highest defect density, for example. Such a concrete prioritization of code classes is covered by selection rules, which operationalize assumptions.

If certain code classes are prioritized, test cases have to be selected or developed, and a focused testing activity can be conducted. Depending on the QA strategy, more or fewer code classes could be selected for the focused testing activity. For example, if the goal is to save effort, only the top-priority code classes would be used for a focused testing activity. If, on the other hand, the goal is to find more defects (i.e., if effectiveness is to be improved), more code classes would be selected. Ideally, both benefits are achieved with such a strategy (i.e., efficiency improvement).

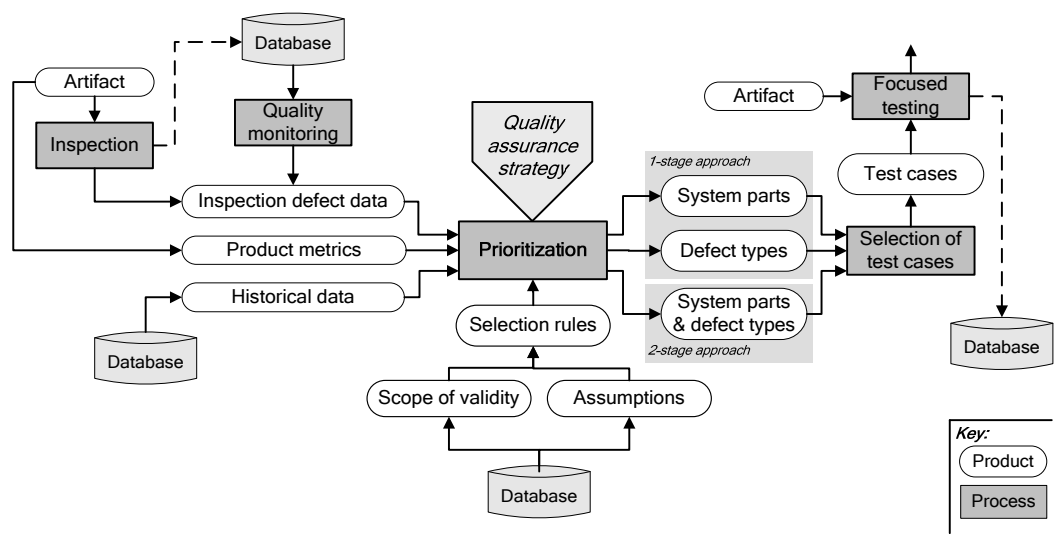

**Figure 1:**    In²Test Approach

In addition, further product metrics and/or historical data can be combined with inspection metrics in order to improve the prioritization. It should be mentioned that assumptions have to be re-validated in each new context; in other words, as-sumptions and concrete selection rules are only valid within a certain scope of context. Finally, in addition to prioritizing system parts, one could also prioritize defect types (but this is not in focus of this article). Figure 1 gives an overview of the concepts of the In²Test approach.

Besides the fact that a combination of different QA activities usually outperforms a single QA activity, three concrete goals that may be addressed with the In²Test approach are: effort reduction, improvement of effectiveness, and improvement of efficiency. Effectiveness is the number of defects found; efficiency is the number of defects found per time unit. These three goals are shown as improvement scenarios in Figure 2.

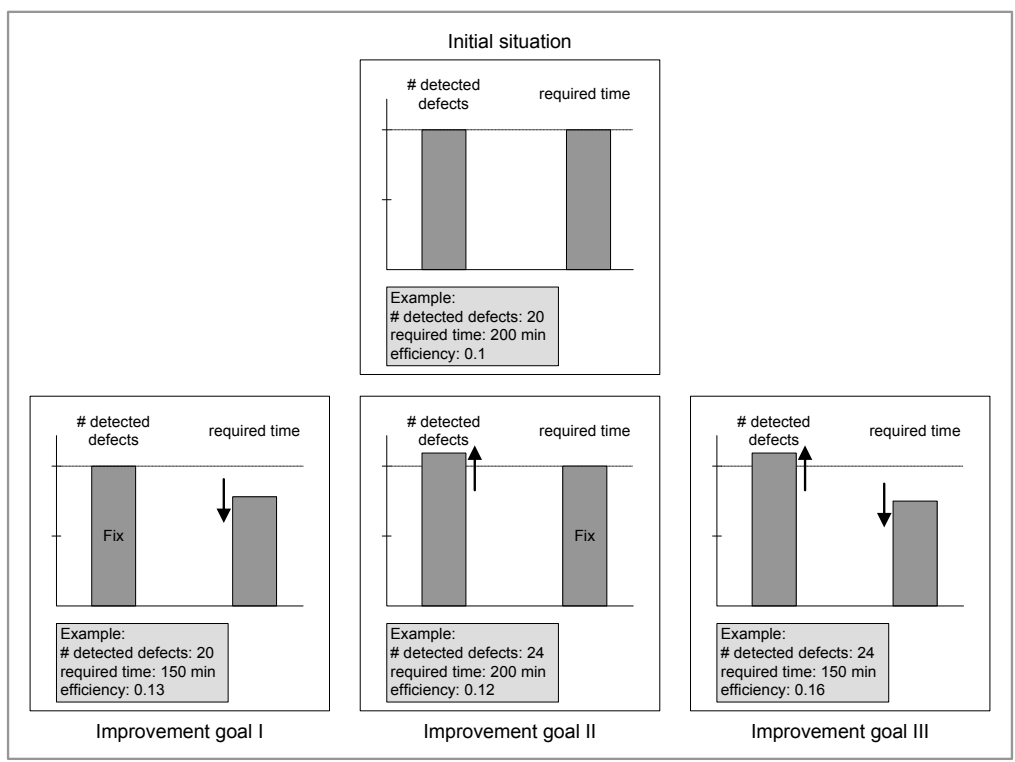

**Figure 2:**    Goal setting for improving QA strategies

Consider the initial situation at the top where effectiveness and efficiency are shown together with concrete exemplary values. The first benefit shows an improved efficiency value, with the same number of defects being found in less time. The second benefit illustrates an improved efficiency value, with the effort value being fixed despite a higher number of defects being found. The third benefit is an improvement of the number of defects found and a reduction in the time consumed. This article, respectively the presented integrated approach, is able to address each of the three mentioned benefits. However, recent evaluations focused on the first improvement goal, and consequently, the results presented in this article will do so as well.

## 3 Evaluation Results and Implications

### 3.1 Main results of two case studies

Two evaluations of the In[2]Test approach were conducted [19-21]. In one evaluation, the main goals were to analyze the applicability of the approach, the improvement in efficiency, and the general study design. A code inspection and testing on the unit and system levels were conducted. The main results were that the approach could be applied and that an improvement in efficiency of up to 30 percent was possible depending on the assumption and selection rule applied. However, one important prerequisite for the application of the approach was that the system under test had to be highly testable. For more details on the design and the results, see [19].

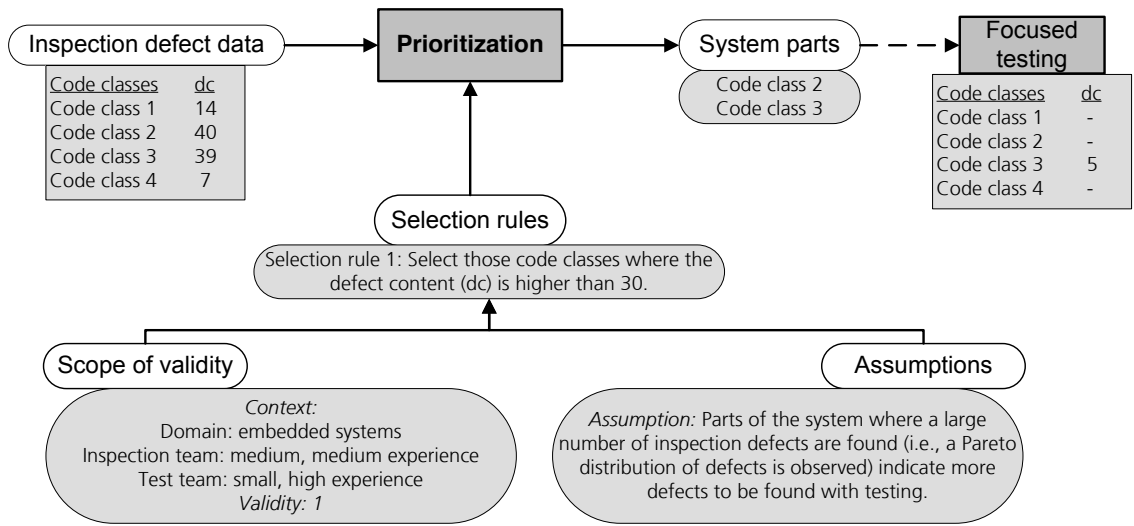

**Figure 3:** Application of In[2]Test

The second evaluation focused on analyzing efficiency improvements and on comparing different assumptions. Figure 3 gives an overview of how the In2Test approach was applied concretely and presents real data from the evaluation. After a code inspection had been done and the quality monitoring proved the results to be valid, a defect profile was derived showing the absolute number of defects (i.e., defect content (dc)) per code class. Next, prioritization was done, using an assumption and a derived selection rule. One assumption used in the case study was a Pareto distribution, i.e., in code classes where a significant number of defects were found during the inspection, more defects were assumed to be found during testing. A selection rule operationalized this assumption as can be seen in Figure 3. In addition, certain context factors were determined and the validity of the selection rule was considered, which was '1' in this case study. The reason is that the assumption (and its derived selection rule) had been proven to be valid once before during an earlier QA run in the same context. Based on this selection rule, two code classes were prioritized (i.e., code classes two and three). A later testing activity showed that five more defects were found in code class 3, which resulted in an effort improvement of about 10 percent with effectiveness remaining the same (remark: each of the four code classes were tested and defect numbers and effort were documented; afterwards, prioritization was done and efficiency improvements were calculated).

A lot of additional inspection metrics (such as defect density) and product metrics (such as complexity and size) were considered and the performance of different assumptions and selection rules was compared. Other assumptions and selection rules applied showed effort improvements of up to 34%, which results in an efficiency improvement of about 50% for testing. A more detailed analysis of about 120 different selection rules indicated that selection rules using inspection results were more efficient in our context than those using traditional size (e.g., lines of code) or complexity metrics (e.g., McCabe complexity). For more details, see [20-21].

## 3.2    Discussion

Both case studies followed a post-analysis design, meaning that testing was conducted first without using the inspection results and an analysis of the efficiency improvements when focusing testing was done afterwards. In order to investigate the approach and to understand the relationships between inspection and testing, this was a reasonable way for evaluations. With respect to an industrial application of the approach, one might start the same way, i.e., analyze historical inspection and testing data, respectively current defect data. However, in order to be able to apply the approach in a pro-active manner, i.e., to prioritize parts and focus testing on these parts, the context needs to be stable, and the assumptions and selection rules need to be validated in that context. Since there are often very large and complex systems, and since there is frequently too little time for testing each part of a system, inspection results might give additional input for focusing on certain parts without overlooking too many defects (especially critical ones).

A lot of different assumptions and selection rules might make sense, and identifying those that lead to the most efficient results requires some effort in the beginning (i.e., analyzing inspection and testing defect data, and deriving assumptions and selection rules); the assumption mentioned in this article and in referring articles [19-21] could be used as a starting point for such analyses. Another idea is to combine selection rules to achieve more appropriate focusing.

In summary, inspections are worth their while when they are applied. An approach that additionally uses inspection results for subsequent test focusing activities might improve their benefit even more. However, due to the size of today's software systems, it is sometimes unreasonable to inspect, for instance, the complete code. In such a case, the defect data from those parts that were inspected can be used to estimate defect numbers in order to undertake appropriate prioritization.

Data from an industrial partner that is currently being analyzed show valuable results for such a scenario.

## 4    Summary and Outlook

In this article, an integrated approach was presented that uses inspection metrics to focus testing activities on those parts of a system that are expected to contain additional defects. In addition, product metrics and historical data are considered and can be combined with inspection metrics. The approach does not replace existing approaches for focusing, but could be used in addition to them, with the aim being more appropriate prioritization of defect-prone parts in order to improve effectiveness, respectively efficiency.

With respect to future work, two main directions could be identified: improvements of the approach and evaluations. For instance, instead of using inspection results, defect information from other types of static analysis might be used. In addition, expert experience might also be used to focus on parts of a system. More fine-grained prioritization (e.g., not only omitting or using code classes completely, but also defining the number of test cases for each class) might lead to a more efficient strategy. Finally, inspection results from design or requirements might improve the focusing activity.

In order to be able to apply the approach in different contexts, more knowledge about the relationships between inspections and testing is necessary. Though a lot of empirical evidence exists with respect to inspection and testing, their integration has not been considered very well yet. Consequently, more evaluations are necessary in order to use the approach effectively and efficiently. Besides focusing on parts of a system, defect types could also be prioritized.